\documentclass[aps,prd,twocolumn,superscriptaddress,showpacs]{revtex4}
\usepackage{graphicx}
\usepackage{dcolumn}
\usepackage{bm}
\usepackage{natbib}
\usepackage{multirow}

\newcommand{\muka}{{\mu{\rm K~arcmin}}}
\newcommand{\rwpe}{w_{{\rm P},{\rm eff}}^{-1/2}}
\newcommand{\rwpesq}{w_{{\rm P},{\rm eff}}^{-1}}

\newcommand{\beq}{\begin{equation}}
\newcommand{\eeq}{\end{equation}}
\newcommand{\beqa}{\begin{eqnarray}}
\newcommand{\eeqa}{\end{eqnarray}}

\begin{document}

\title{$\mathbf{B}$ polarization of cosmic microwave background as a tracer of strings} 
\author{Uro\v{s} Seljak}
\affiliation{Department of Physics, Princeton University, Princeton NJ 08544, U.S.A.}
\affiliation{International Center for Theoretical Physics, Trieste, Italy}
\author{An\v{z}e Slosar} 
\affiliation{Faculty of Mathematics and Physics, University of Ljubljana, Slovenia}

\date{\today}


\date{\today}

\begin{abstract}
  
  String models can produce successful inflationary scenarios in the
  context of brane collisions and in many of these models cosmic
  strings may also be produced. In scenarios such as KKLMMT the string
  contribution is naturally predicted to be well below the
  inflationary signal for cosmic microwave background (CMB)
  temperature anisotropies, in agreement
  with the existing limits.  We find that for $B$ type polarization of
  CMB the situation is reversed and the dominant signal comes from
  vector modes generated by cosmic strings, which exceeds the gravity
  wave signal from both inflation and strings.  The signal can be
  detected for a broad range of parameter space: future polarization
  experiments may be able to detect the string signal down to the
  string tension $G\mu=10^{-9}$, although foregrounds and lensing are
  likely to worsen these limits.  We argue that the optimal scale to
  search for the string signature is at $\ell\sim 1000$, but in models
  with high optical depth the signal from reionization peak at large
  scales is also significant. The shape of the power spectrum 
  allows one to distinguish the string signature from the gravity
  waves from inflation, but only with a sufficiently high angular
  resolution experiment.

\end{abstract}

\pacs{98.80.Jk, 98.80.Cq}

\maketitle

\setcounter{footnote}{0}

\section{Introduction}

Inflation is a theory that predicts the universe has undergone a
period of exponential expansion sometime in the early epoch of its
history
\cite{1981PhRvD..23..347G,1981MNRAS.195..467S,1982PhLB..117..175S,1982PhRvL..48.1220A}.
The success of inflation is due to the fact that it solves a number of
problems in cosmology, such as flatness and horizon. Even more
importantly, it can explain the origin of structure formation in the
universe, as quantum fluctuations are stretched to cosmological scales
during the exponential expansion
\cite{1981JETPL..33..532M,1982PhRvL..49.1110G,1983PhRvD..28..679B,1982PhLB..115..295H,1982PhLB..117..175S}.
Inflation makes a number of potentially observable predictions, such
as a nearly scale invariant shape of the primordial spectrum,
adiabatic nature of perturbations, absence of detectable nongaussianty
and zero curvature.  It has passed all of these observational tests so
far and is the leading paradigm for the origin of structure formation
in the universe.

One of the most distinguished tests of inflation is its prediction of
gravity waves, which are generically produced in all models of
inflation and reflect the energy scale of inflation. No gravity waves
have been detected so far, but the current limits are weak. This is because
the gravity wave signal is expected to contribute only on large
angular scales, where cosmic variance limits the precision of their
extraction. With cosmic microwave background (CMB)
temperature spectrum the limits cannot be improved much better than the
existing limits.  It is often argued that a smoking gun for their
detection is $B$ type polarization of CMB, which is not contaminated
by scalar perturbations and thus not limited by cosmic variance
\cite{1997ApJ...482....6S,1997PhRvD..55.1830Z,1997PhRvD..55.7368K}. It
is only limited by detector noise and other contaminants such as
foregrounds and weak lensing which converts $E$ polarization into $B$.
The importance of this probe and its promise for a ground breaking
discovery has been recognized by the wider community and there are
many ground based CMB polarization experiments in various stages of
planning or building. Moreover, a future satellite mission dedicated
to $B$ type polarization has been identified as one of the NASA
Einstein probes to be built over the next decade, although recent
budgetary constraints may delay its implementation.

In contrast to the success of inflation as a phenomenological model,
producing inflation from fundamental theories like string theory has
been more of a challenge.  There has been recent progress on this
subject in the context of the brane world scenarios, which suggest
that we may be living on a hypersurface embedded in higher dimensions.
Brane inflation is a generic outcome of scenarios where branes collide
and heat the universe, initiating the hot big bang
\cite{1999PhLB..450...72D}. Another generic prediction of these models
is that cosmic strings are produced during the brane collision
\cite{Sarangi:2002yt}.

Cosmic strings and other topological defects have long been one of the
candidates for the origin of structure formation, but this scenario
has been shown to lead to predictions incompatible with observations
such as the power spectrum of cosmic microwave background 
temperature anisotropies
\cite{1997PhRvL..79.1611P,Allen:1997ag,Albrecht:1997mz}.  In the
absence of explicit predictions it seems unnatural to have topological
defects play a subdominant, yet non-negligible role, so such models
have largely been abandoned.  This has changed in the context of
string inspired models of brane inflation, some of which naturally
explain why cosmic strings have a small contribution to the CMB
relative to inflation, therefore accomodating the observational
limits.  This has led to a significant revival of all aspects of
cosmic string scenario, including new theoretical motivations,
phenomenological implications and direct observational searches.
While not all string inspired models of inflation produce significant
cosmic strings \cite{Dvali:2003zj,Blanco-Pillado:2004ns}, 
the possibility of having an
observable window to the string theory is too important to be ignored.

One of the most fully developed models of string inflation is KKLMMT
model, in which $\bar{D3}$ brane is sitting at the bottom of a throat
and a $D3$ brane is moving towards it until they collide
\cite{Kachru:2003sx}.  Brane inflation in this model leads to an
adiabatic spectrum of fluctuations and can satisfy all the
observational constraints, provided that the inflationary potential is
sufficiently flat, which requires fine-tuning at a few percent level.
In the collision a network of effectively local strings is generated
and these may remain stable due to the warping of the compact
dimensions. The tension of these strings is naturally predicted to be
small and their contribution to CMB is well below the current limits
\cite{Copeland:2003bj}. 
Since cosmic variance limits our ability to
distinguish between the two components it is likely that we cannot
detect their signal in CMB temperature anisotropies, unless the level
is just below the current limits.

The above arguments suggests that one cannot observe a small signal
from cosmic strings in CMB because it is dominated by the scalar
perturbations from inflation. However, just as in the case for gravity
wave signal from inflation, the situation changes if one considers $B$
type polarization of CMB, which does not receive a contribution from
primordial scalar modes apart from what is produced by gravitational
lensing \cite{1998PhRvD..58b3003Z}. First calculations of CMB
polarization in global cosmic strings and other global defects have
found that $B$ polarization signal can be significant and is dominated
by vector modes \cite{Seljak:1997ii}, but there has been some
controversy on the applicability of these results to local strings
relevant for the string inspired models considered here
\cite{Albrecht:1997mz,Wyman:2005tu,Fraisse:2006xc}.  The purpose of
this paper is to explore the predictions of cosmic strings in
polarization in the context of string inflation models and to
establish prospects for their future detectability.  While we work in
a specific context of KKLMMT model, our results on the detectability
levels in terms of string tension are more general and applicable not
only to other string inspired models of inflation, but also to the
more standard cosmic strings production scenarios such as those based
on (SUSY) GUT scale phase transition \cite{Jeannerot:2003qv}.

\section{Model and Analysis Method}

In this paper we adopt a simple generalization of KKLMMT scenario
considered in \cite{Firouzjahi:2005dh}. The potential in this model is
of the form
\begin{equation}
V={1 \over 2}\beta H^2 \phi^2 + V_0\left(1-{A \over \phi^4}\right),
\label{eq1}
\end{equation}
where $H$ is the Hubble parameter and $\phi$ is the scalar field
representing the separation between the branes. Second term in
equation \ref{eq1} represents the potential of the branes when they
are far apart and is driving inflation, third term represents
attractive potential during collision which causes inflation to end
aat collision, while first term is conformal coupling like and arises
from additional contributions such as from K\" ahler potential and is
spoiling the slow-roll conditions, unless $\beta$ is sufficiently
small, which requires fine tuning. It also has to be positive to
prevent the brane repelling each other even before the collision takes
place.  For observationally relevant range with $0<\beta<0.05$ this
model gives, very approximately, for primordial spectral index $n_s
\sim 0.98+\beta$, for tensor to scalar ratio $\log r \sim -8.8+60
\beta$ and for string tension $\log G\mu \sim -9.4+30\beta$. Gravity
wave signal from inflation is extremely small in these models and is
not expected to be detected for any relevant value of $\beta$.
Adopting the existing constraints on the slope of spectral index in
the absence of tensors and running, $n_s<1.03$ (95\% c.l.) requires
$\beta <0.05$. However, the latest WMAP constraints from their 3 year
analysis are even more stringent and only a narrow range of parameter
space for this model is still allowed \cite{2006astro.ph..3449S}.
Nevertheless, even if the current model is ruled out we expect that
the more generic predictions presented here will survive.
Specifically, in this model the predicted string tension $G\mu$ is
well below the existing limits from CMB temperature, which are around
$2.7\times 10^{-7}$ \cite{Wyman:2005tu,Fraisse:2006xc,pogoerr}, yet the tension
is also larger than $3\times 10^{-10}$ for all of the parameter space.
An analysis of recent cosmic microwave background data, large scale
structure data, supernovae Ia and Lyman-$\alpha$ forest data results
in an upper limit of $G\mu<2.3\times 10^{-7}$ at 95\% confidence
limits for a fixed fiducial cosmic string model \footnote{McDonald,
  Seljak \& Slosar, \emph{in preparation.}}.

To describe the effects of strings on the structure formation one must
first solve for their evolution given their initial conditions of a
string network.  As the universe expands new strings continously enter
the horizon, intersect and develop loops, which then decay away
through radiation of gravity waves and possibly other fields.  This
reconnection probability can be much smaller than one, which is one of
the distinguishing new features of cosmic strings produced by
fundamental strings as opposed to those based on field-theory driven
symmetry breaking.  One expects the string network to achieve scaling
both in matter and radiation dominated epochs, so that the network is
self-similar relative to the horizon scale.  Evolution of this string
network is nonlinear and has to be modelled numerically.  Since most
of the small scale smoothing comes from small loops and wiggles a
large dynamic range is required and the convergence of the simulations
has been difficult to achieve.  Results from recent simulations in an
expanding universe suggest that the convergence has been achieved, but
it is not clear whether the results from different groups are entirely
in agreement \cite{Ringeval:2005kr,Vanchurin:2005pa}.  Generally,
while there is considerable uncertainty in the evolution of string
network on small scales, the situation is more robust on horizon
scales where causality plays an important role \cite{Turok:1997gj}.

In this work we use the public cosmic string code developed and
maintained by L. Pogosian \cite{Pogosian:1999np}, which has been
calibrated to reproduce the correlation functions of full-scale
simulations.  In \cite{Pogosian:1999np} it was found that the
simulations exhibit a significant amount of string wiggliness at the
resolution scale.  Since small scale simulations are still poorly
resolved one parametrizes the uncertainty in the level of wiggliness
with a free parameter. Increased wiggliness of the strings can be
accounted for by modifying the string energy-momentum tensor
\cite{Pogosian:1999np}. 
Cosmic string network acts as a continuous source of metric
perturbations.  To compute their effect on CMB and large scale
structure (LSS) one needs to know the unequal-time correlators of
energy-momentum tensor.  We use Pogosian's code \cite{Pogosian:1999np}
for computing the cosmic string energy momentum source terms, which
are then fed into a Boltzmann code.  Pogosian's code uses CMBFAST
\cite{1996ApJ...469..437S} and we wrote a separate code using CAMB
\cite{2000ApJ...538..473L} to verify the calculations, which now agree
with each other \cite{pogoerr}.  For the purpose of this paper it is
particularly relevant to know the relation between scalar, vector and
tensor sources. All three add incoherently to the CMB temperature
perturbations, but only vector and tensor modes contribute to $B$ type
polarization.  Since the perturbations themselves are incoherent they
result in a broad peak in CMB power spectrum, contrary to coherent
oscillations seen in the data and in theoretical predictions of
inflation.  As a result, cosmic strings can only contribute up to 10\%
of the total contribution to CMB on observed scales and this
translates in the upper limit on the dimensionless string tension
$G\mu<2.7\times 10^{-7}$ \cite{Wyman:2005tu,Fraisse:2006xc,pogoerr}.  Anything
below this is allowed by the current data and due to cosmic variance
it will be difficult to improve these limits much in the future using
CMB temperature information only. However, since $B$ polarization only
receives contributions from vectors and scalars the cosmic variance
problem is aleviated: the main contamination to cosmic string signal
comes from gravity waves from inflation and from lensing of $E$
polarization.

Analytical calculations in \cite{Turok:1997gj} have shown there exists
a relation between scalar, vector and tensor perturbations on horizon
scale, which can be used to predict the corresponding fractions of the
three components in the CMB, subject to some important assumptions
such as comparable correlation length. This prediction was shown to be
reasonably well satisfied in the simulations of global strings
\cite{Turok:1997gj}.  One of the outcome of these calculations is that
vector modes play an important role and may, depending on the model,
even be the dominant source of perturbations.  In particular, they
were shown to dominate over the tensor modes
\cite{1997PhRvL..79.1611P,Seljak:1997ii} in global string simulations.
However, it has been often argued that the evolution of local strings
may be significantly different from that of global strings so that
insights attained in the global case case may not apply to the local
case, so a more direct calculation of local strings is needed.

\section{Results}

\begin{figure*}[htbp]
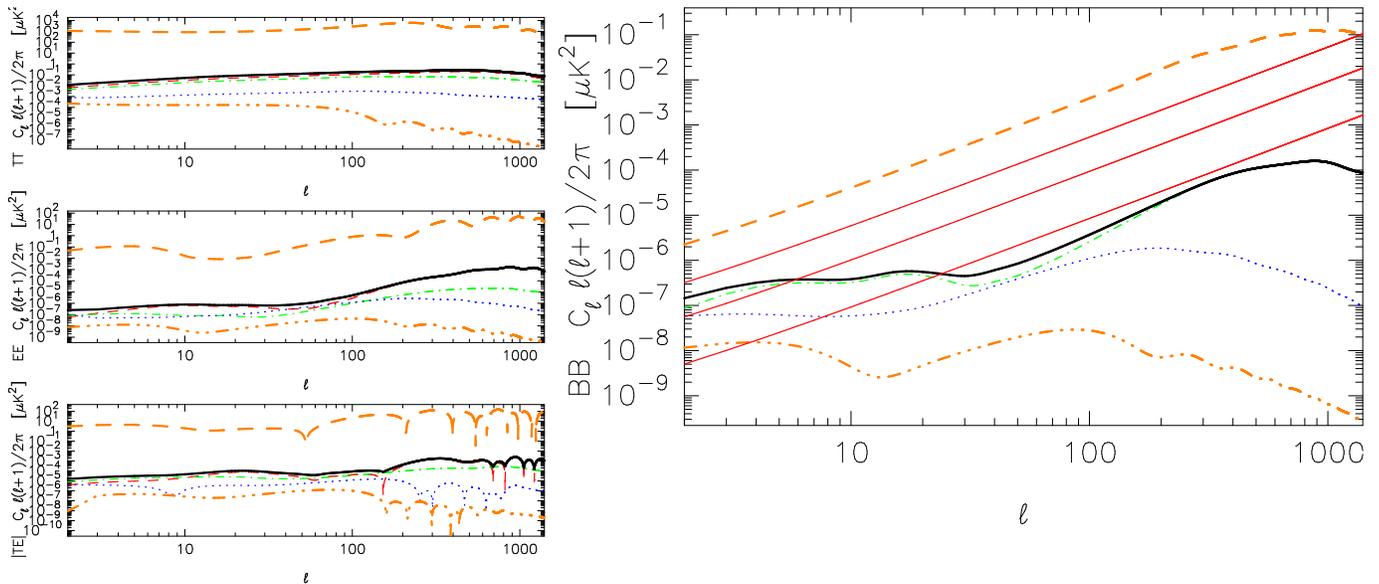

  \centering
  \begin{tabular}{cc}
    \includegraphics[height=0.4\linewidth, angle=-90]{s1nt.eps} &
    \includegraphics[height=0.6\linewidth, angle=-90]{xs1nt.eps} \\
  \end{tabular}
  \caption{This figure shows various CMB temperature and polarization
    power spectra. The left panel shows $TT$ (top), $EE$ (middle) and
    the absolute value of the $TE$ (bottom) power spectra, while the
    larger right-hand side plot shows the same for the $BB$ power
    spectrum. The thick dashed and dot-dashed lines (orange) correspond to the
    inflationary 
    contribution for the scalar and tensor modes, respectively; in the
    $BB$ power spectrum the scalar contribution comes exclusively
    from lensing of the $EE$ polarisation modes by the intervening
    structure between us and the surface of the last scattering. The
    thin dashed (red), dot-dashed (green) and dotted (blue) lines show
    the scalar, vector and tensor contributions to the total string
    contribution plotted as a thick solid line (black), assuming 
$G\mu = 10^{-8}$. The red thin
    straight lines in the BB power spectrum correspond to rough limits
    on residual noise
    obtained by cleaning the lensing contamination from $E$ polarization
    by the quadratic estimator (top) and iterative
    method (middle), while the bottom is
    the instrumental noise for a hypothetical future instrument
    with polarisation sensitivity of $\sim 0.25 \mu$K arcmin and beam size of
    $2'$.}
  \label{fig1}
\end{figure*}

Our calculations of CMB predictions using modified Pogosian's code are shown in
Figure \ref{fig1}, assuming $\beta=0.02$ and smooth strings with no
extra small scale wiggliness.  In this example we have $r=10^{-6}$ and
$G\mu=10^{-8}$.  In temperature, $E$ polarization auto-spectra and
their cross-correlation the string signal is orders of magnitude below
the scalar contribution from inflation and for much of the parameter
space strings cannot be detected \footnote{Note that the string signal
  may be detectable on much smaller scales where diffusion damping
  suppresses anisotropies from recombination epoch, but not integrated
  Sachs-Wolfe anisotropies generated along the line of sight by cosmic
  strings.  This will be addressed in a future publication.}.

The situation is reversed for $B$ polarization, where there is no
primordial scalar contribution from inflation. Moreover, the tensor
contribution from inflation is well below the string contribution.
This is true for all the values of $\beta$ in this model.  It is
interesting to note that over most of the range vector perturbations
exceed tensor perturbations from strings, just as it was found for
global strings \cite{Seljak:1997ii}.  The exception may be around $\ell
\sim 30-100$, where however the signal is weakest compared to the
noise.  As Figure \ref{fig1} shows there are two peaks in the signal,
reionization peak for $\ell<20-50$ and the recombination peak at $\ell \sim
1000$.  Reionization peak has the origin in the large quadrupole
moment on the scale of horizon due to the free streaming of photons
after recombination.  Because of reionization these photons are
rescattered and this generates polarization on the angular scale of
horizon at that epoch.  The reionization peak amplitude strongly
depends on the adopted value of optical depth $\tau$ and increases by
an order of magnitude between $\tau=0.04$ and $\tau=0.17$.  Recent WMAP
results \cite{2006astro.ph..3449S} have converged onto $\tau\sim 0.1$,
so we adopt this value in our calculations.  The main
recombination peak is dominated by incoherent contributions generated
by cosmic strings during and after recombination.  It is much less
dependent on reionization optical depth.  Overall, we find very
similar results to the global string predictions in
\cite{1997PhRvL..79.1611P,Seljak:1997ii}, suggesting that there is
little qualitative difference between global and local strings in
their CMB predictions, at least for the smooth string models
considered in this example.

\begin{table*}
\begin{tabular}{cc|cccc}

\hspace*{1cm}& \hspace*{1.8cm}  &\hspace*{1cm}    &
       \multicolumn{2}{c}{$\left(\sigma_{(G\mu)^2}\right)^{1/2}/\ 10^{-9}$}
       &\hspace*{1.8cm} \\

       &          & \multicolumn{2}{c}{ $\rwpe = 0.8\muka$}
     & \multicolumn{2}{c}{ $\rwpesq=C_{\rm scalar,lensed}(\ell)$ } \\

 $\tau$ & $\alpha$   & $\ell<100$ & $100<\ell<1200$ \hspace*{0.5cm} &
        \hspace*{0.5cm} $\ell<100$ &  $100<\ell<1200$ \\
\hline
0.04 & 1.0 & 3.4 &   1.4 &    22.9&    9.3 \\
0.04 & 1.9 &  5.3 &   1.8 &    35.9&   12.4 \\
0.04 & 3.0 &  7.6 &   2.4 &    51.6&   16.5 \\
\hline
0.10 & 1.0 &  2.2 &   1.4 &    14.0&    9.3 \\
0.10 & 1.9 & 3.3 &   1.9 &    20.6&   12.4 \\
0.10 & 3.0 & 4.5 &   2.6 &    28.6&   16.5 \\
\hline
0.17 & 1.0 &  1.5 &   1.5 &     8.9&    9.3 \\
0.17 & 1.9 &  2.2 &   2.1 &    13.0&   12.4 \\
0.17 & 3.0 &   3.0 &   2.7 &    18.2&   16.5 \\
\hline
\end{tabular}
\caption{This table show the detectability limits for $G\mu$ for
        various combinations of the optical depth ($\tau$), wiggliness
        ($\alpha$), $\ell$ range and assumed noise.}

\label{tab1}
\end{table*}

At what level can the string signal be detected in CMB?  As discussed
above, for CMB temperature anisotropy $T$ the cosmic variance prevents
one to detect the string contribution if the signal is below a few
percent of the inflationary signal and if only $\ell<1000-2000$
information is used.  The same conclusion is valid for $E$
polarization and its cross-correlation with $T$.  Thus it is the $B$
polarization auto-correlation that offers the best prospects for a
detection given sufficiently high signal to noise detector, since it
is not contaminated by primary scalar modes.  However, even $B$
polarization is contaminated, because gravitational lensing converts
some of the scalar $E$ polarization into $B$
\cite{1998PhRvD..58b3003Z}.  The dashed curve at the top of bottom
right panel of Figure \ref{fig1} shows this lensing induced $B$
polarization generated from $E$ polarization as computed by CMBFAST
\cite{1996ApJ...469..437S}.  It has roughly a white noise power
spectrum up to $\ell \sim 700$, beyond which it gradually flattens and
eventually drops.

This lensing induced $B$ polarization can be reduced if one has
information on the projected lensing potential, which allows one to
delens the CMB
\cite{2002PhRvL..89a1304K,2002PhRvL..89a1303K,2004PhRvD..69d3005S}.
This can be achieved using the nongaussian correlations in the CMB
temperature or polarization
\cite{2002ApJ...574..566H,2003PhRvD..67d3001H,2003PhRvD..68h3002H} or
from external information obtained from other tracers (e.g.  21 cm
fluctuations \cite{2004NewA....9..417P,2005PhRvL..95u1303S}).  If one
assumes quadratic estimator from \cite{2002ApJ...574..566H} then one
can reduce the lensing noise in the white noise regime by a factor of
7. Iterative estimator can in cases of very low detector noise, such
as a hypothetical CMBPOL type satellite, give $\rwpe=0.8\muka$ for a 2'
beam, i.e. an improvement of a factor of 40 relative to the no lens
noise cleaning \cite{2004PhRvD..69d3005S}.  This is still several
times above the instrument noise, so lensing noise always dominates.
All three noise curves are shown in Figure \ref{fig1}.  We have
adopted the white noise approximation for quadratic and iterative lens
cleaning, although in practice the situation may be better since the
lensing noise itself decreases below the white noise on small scales.
On the other hand, it is unclear whether the iterative method can
obtain this reduction on small scales in this model, since the method
works on the assumption that there is no $B$ signal except that coming
from lensing and so it must be generalized to account for the signal
from strings.

For a given signal and lensing induced $B$-mode noise power spectrum the
resulting uncertainty on $G\mu$ is:
\begin{equation}
\sigma_{(G\mu)^2}^{-2} =
f_{\rm sky}\sum_\ell {2\ell+1 \over 2}
w_{\rm{P},\rm{eff},\ell}^{2}  
\left({C_\ell^{BB} \over (G\mu)^2}\right)^2,
\label{eq:error}
\end{equation}
where $C_{\ell}^{BB}$ is the string power spectrum of $B$ modes as in
Figure~\ref{fig1}, and $w_{\rm{P}, \rm{eff},\ell}=C^{BB}_\ell({\rm residual})$
is the inverse noise variance per solid angle per polarization that
has units of $(\muka)^2$ and represents the noise from combined
instrument noise and lensing residuals that limits the detectability
of the signal.

It is worth considering the reionization peak and the main peak
separately.  The reionization peak depends sensitively on the Thomson
scattering optical depth $\tau$ due to reionization, which is still
somewhat uncertain, while the main peak sensitivity is much weaker.  
In addition,
incomplete sky coverage and foregrounds are particularly worrisome on
large scales, so the reionization peak may be more difficult to
observe than the recombination peak
\cite{2005PhRvD..72l3006A,2006JCAP...01..019V}.  The results of the
calculations for various levels of wiggliness are given in Table
\ref{tab1}.  For reionization peak, using information with $\ell<100$, we
find that the error on $G\mu$ varies between $1.5 \times 10^{-9}$ for
$\tau=0.17$ and no wiggliness to $7.6 \times 10^{-9}$ for $\tau=0.04$
and high wiggliness, assuming noise levels of iterative lens cleaning
procedure in a CMBPOL type experiment. In all cases full sky is
assumed.  In the other extreme we assume full lensing noise with no
delensing. In this case we find the
limits are between $10^{-8}$ and $5 \times 10^{-8}$ on $G \mu$.

For partial-sky coverage, Eq.~(\ref{eq:error}) must be modified to
take into account sky cuts; while $\sigma_{(G\mu)^2}\propto
f_{\rm sky}^{-1/2}$ for the small scale peak on sub-degree scales, the
reionization peak present at $\ell<20$ exhibits a much more complicated
dependence on the survey geometry due to cross-leakage of $E$ and $B$
modes induced by, e.g. the Galactic Plane cut
\cite{2002PhRvD..65b3505L,2003PhRvD..67b3501B}.  In
\cite{2005PhRvD..72l3006A} it was argued that the scaling with sky
fraction becomes $\sigma_{(G\mu)^2}\propto f_{sky}^{-2}$ for 
$f_{sky}>0.7$. This could
further weaken the limits from large scales given in Table \ref{tab1}.
The degradation is worst for models with late reionization because
this pushes the $B$ reionization peak to the lower multipoles where
sky-cut effects are most severe.  In particular, using the analysis in
\cite{2005PhRvD..72l3006A} and assuming we can remove dust foregrounds
at the 0.01\% level of unpolarized emission, we find that $G \mu \sim
10^{-8}$ can be achieved from large scales, but that it will be
difficult to go below that using reionization peak information alone.

The situation is better for the main recombination peak at $\ell \sim
1000$. We find that we can achieve between $G\mu =1.4 \times 10^{-9}$
and $2.7 \times 10^{-9}$ depending on the wiggliness and assuming
iterative delensing procedure with high angular resolution and low
noise detector like CMBPOL.  For the case of no lens cleaning the
numbers vary between $9 \times 10^{-9}$ and $1.6 \times 10^{-8}$. Even
in this case the improvements are at least one order of magnitude
better than the current limits.  While the level of polarization
foregrounds is poorly known on these scales, galactic emission tends
to be smooth and decreases towards smaller angular scales, although
this prediction could be modified if there are small scale magnetic
fields in our galaxy generating small scale power in synchrotron
polarization.

We should warn that there is still considerable uncertainty regarding the
predictions of string models and our results on the value of $G\mu$
should be viewed as qualitative. However, both the shape of the power
spectrum and string tension normalization appear to be very similar
among different groups, suggesting that the remaining uncertainties
may not make much of a quantitative difference to the results found
here.  On the other hand, varying the details of the string network evolution,
such as decay rate, intercommutation probability and wiggliness, can
change the coherence length and move the peak of the power spectrum in
CMB temperature and polarization.  It can also change the ratios of
scalar to vector to tensor contributions to CMB temperature or
polarization.

\section{Discussion}

It has long been recognized that gravity waves are
a natural outcome of inflation with an amplitude in CMB that may be close
to observed limits and may be best observed in CMB $B$ polarization 
experiments. 
Recent models of inflation in brane 
collisions suggest a 
very similar situation for cosmic strings generated from the brane collisions
and a natural 
outcome of such models may also be CMB polarization 
at a detectable level in $B$ channel. 
We find the string signal is dominated by vector modes over 
tensor modes and, for 
models analyzed here, both of these exceed  
the gravity wave signal from inflation. 

Thus string inspired models of inflation may challenge the 
conventional view that a detection of
$B$ type polarization in cosmic microwave background (CMB)
will demonstrate the existence of gravity waves 
in the early universe and measure the energy scale of inflation. 
If only the large scale signal is observed 
in $B$ polarization 
then it may be difficult to 
distinguish between  
the string and inflation scenarios.  
This is because there is only a finite number of modes being observed 
and cosmic variance prevents one from accurately determining the shape 
of the power spectrum on large scales, and the differences 
between the two models are small (Figure \ref{fig1}). 

On the other hand, if high angular resolution is available, then
separating cosmic string signal from inflationary gravity wave signal
should be possible using the power spectrum shape informations, since
strings predict the signal dominates at $l \sim 1000$, while gravity
wave signal from inflation peaks at $l \sim 100$ and decays away on
smaller scales.  For our most optimistic case we find that string
tension down to $G\mu \sim 10^{-9}$ can be detected with
$B$-polarization, two orders of magnitude below the current limits.
Galactic foregrounds and gravitational lensing may considerably weaken
these limits and $G\mu \sim 10^{-8}$ may be a more realistic target.
These limits are in the range of current model predictions, although
they do not cover the entire range since even lower values of string
tension are possible.  Nevertheless, searching for this signature
provides additional motivation for upocoming CMB polarization
experiments, specially on small angular scales, where only the lensing
induced $B$ polarization was previously expected to be seen. A search
for excess signal on these scales may instead reveal a signature of
string physics.

\section*{Acknowledgements}
We thank Levon Pogosian for help with his string code and Antony Lewis
for very useful help in clarifying various aspects of vector
perturbations evolution in CAMB \cite{2000ApJ...538..473L,
  2004PhRvD..70d3518L,Lewis:2004ef}. U.S. is supported by the Packard
Foundation, NASA NAG5-1993 and NSF CAREER-0132953. AS is supported by
the Slovenian Research Agency grant Z1-6657.

\bibliography{cosmo,cosmo_preprints}
\end{document}